\documentclass[12pt]{article}
\RequirePackage{ifthen}

\newboolean{shortversion}
\setboolean{shortversion}{false}

\newboolean{slacreport}
\setboolean{slacreport}{true}

\newboolean{elsevier}
\setboolean{elsevier}{false}

\RequirePackage[dvips]{graphicx}

\newcommand{\progname}{{\tt TraFiC${}^{\texttt{4}}$}}
\newcommand{\Lover}{L_{OT}}

\newcommand{\papertitle}{Numerical Calculation of Coherent Synchrotron
  Radiation Effects Using \progname}

\RequirePackage{andreas}

\begin{document}

\ifthenelse{\boolean{elsevier}}{\journal{NIMA}\date{}}{}

\ifthenelse{\boolean{slacreport}}{
\renewenvironment{abstract}
{\vfill\begin{center}{\bf\large Abstract}\end{center}\begin{quote}}
{\end{quote}\vfill}}{}

\ifthenelse{\boolean{elsevier}}{
\begin{frontmatter}
\title{\papertitle}
\author[SLAC]{A.~Kabel\thanksref{CORR}}
\author[DESY]{M.~Dohlus}
\author[DESY]{T.~Limberg}
\address[SLAC]{\slacaddress}
\address[DESY]{\desyaddress}
\thanks[CORR]{Corresponding author; E-Mail: \tt andreas.kabel@slac.stanford.edu}}{}

\ifthenelse{\boolean{slacreport}}{
\thispagestyle{empty}
\renewcommand{\thefootnote}{\fnsymbol{footnote}}
\begin{flushright}
{\small
SLAC--PUB--8559\\
August 2000\\}
\end{flushright}
\vspace{.8cm}
\begin{center}
{\bf\large\papertitle\footnote{Work supported by
Department of Energy contract DE--AC03--76SF00515.}}\\
\vspace{1cm}
A.~Kabel\\ \slacaddress\\
\medskip
M.~Dohlus, T.~Limberg\\ \desyaddress
\end{center}
\pagestyle{plain}
}{}

\begin{abstract}
  Coherent synchrotron radiation (CSR) occurs when short bunches travel
  on strongly bent trajectories. Its effects on
  high-quality beams can be severe and are well understood qualitatively. 
  For quantitative results, however, one has to rely on numerical methods. There
  exist several simulation codes utilizing different approaches. We
  describe in some detail the code \progname{}  developed at DESY for
  design and analysis purposes, which approaches the problem from first
principles and solves the equations of motion either perturbatively or
self-consistently. We present some calculational results
  and comparison with experimental data. Also, we give examples of how
  the code can be used to design beamlines with minimal emittance growth
due to CSR.
\end{abstract}

\ifthenelse{\boolean{slacreport}}{
\begin{center} 
{\it Invited talk presented at the\\International Symposium on New Visions in Laser-Beam
Interactions\\
Tokyo Metropolitan University, Tokyo, Japan \\
October 11th--October 15th, 1999}\\
{\it Submitted to Nuclear Instruments and Methods A}
\end{center}
\newpage
\normalsize
}{}

\ifthenelse{\boolean{elsevier}}{
\end{frontmatter}
}{}

% intr CSR abbr
% intr Slice emittance, sampling/generating particles
% subbunches->bunchlets
% intr CTF2 experiment
% intr slice emittance

\section{Introduction}

\subsection{Effects of Retarded Fields on a Bend}
When short bunches travel on bent trajectories, electromagnetic fields
emitted by the tail  of the bunch may  overtake the head,  influencing
the collective behavior of  the  bunch.   As opposed  to  space-charge
effects, this effect  is, for relativistic   beams, completely due  to
geometry  and thus independent  of energy. On  a bend with radius $R$,
the arc length needed to allow fields to  catch up a distance $\sigma$
is,  to first nontrivial order   in $L/R$, $\Lover = \sqrt[3]{24\sigma
  R^2}$ \cite{Derbenev:1995eg}. The  bunch  then interacts  collectively with
itself; this mechanism is commonly referred to as coherent synchrotron
radiation (CSR).

The longitudinal electric field will cause the bunch to develop a
longitudinal energy gradient, which will lead to a growth of the
projected emittance due to dispersion mismatch, as the energy gains
and losses are induced in regions with non-vanishing dispersion.  For
Free-Electron Laser operation, the interesting quantity is the slice
emittance, i.~e., the emittance of a longitudinally small sub-ensemble
of the bunch.  This quantity may also suffer degradation, as a slice
can pick up an energy spread due to both the transversal and -- in
case the slice tilts due to the beamline geometry -- longitudinal
variation of the CSR fields.  The transverse fields will exert
transverse kicks on the particles, leading to both projected and slice
emittance growth, although this effect is of less importance for the
cases studied so far.  The transverse and longitudinal part of the CSR
force are intricately related to each other, leading to a cancellation
of part of the transverse force by the effective force due to the
energy change in dispersive regions
\cite{Talman:1986is,Lee:1988xx,Derbenev:1996ub}.  Using the usual
textbook formulae for retarded fields, one easily sees that, on a
curved trajectory, the transverse electric field caused by
acceleration is acting instantaneously, while the longitudinal field
needs some build-up length (of the order of $\Lover$) when entering or
a bent region before reaching a steady-state value. The same is true
for the decay to 0 after leaving a bent region. Consequently, CSR
effects are not expected to play a role if $L_{bends} \ll \Lover$.  If
$L_{bends} \gg \Lover$, the system can be treated using analytic
formulae derived for the steady-state case\cite{Derbenev:1995eg}.
When evaluating systems in which $\Lover \approx L_{bends}$,
steady-state formulae are of limited utility, as they disregard the
transient effects.

\section{The Code \progname}
A way to handle this difficulty is a tracking code incorporating a
field calculation from first principles, i.~e., by numerically solving Maxwell's
equations for a bunch of particles traveling through a given beamline.
This can be done by using the method of retarded fields. \progname{} 
(standing for ``\underline{Tra}cking particles in the
\underline{Fi}elds of \underline{C}ontinuous \underline{C}harges in
\underline{C}artesian \underline{C}oordinates'') is a tracking code
implementing this concept.

\subsection{Cartesian Tracking}
The effects  of CSR on  the  bunch described  above  are nonlocal both
temporally  and    spatially:  the   equations   of     motion    are
integro-differential equations.  Thus, the history of  the beam has to
be known to  determine the fields acting  on it.  To store the history
of each particle, using a  global Cartesian coordinate system  instead
of the usual local co-moving one is an obvious choice.
Since a complete history is impossible to keep in memory, the problem
is discretized. The beamline is divided into slices; on each slice,
the relevant parameters of a particle entering the slice are stored by
a first, ``macro-tracking'' run.  These could, in principle, just be
the usual six local phasespace
coordinates\ifthenelse{\boolean{shortversion}}{}{, which are the
  complete set of initial conditions for the local differential
  equations of motion}.  Instead, we use as six independent parameters
the positional vector $\vec q$, the normalized tangential vector $\vec
e = \ta s\vec q(s)$, and the relativistic factor
$\gamma$\ifthenelse{\boolean{shortversion}}{}{ (using the velocity
  vector $\vec\beta$ would not make sense from a numerical point of
  view)}.  We also store the arc length $s$ and time $t$ traveled by the
particle from the zero-point of the trajectory. In these coordinates,
the equations of motion read:
\begin{math}\label{eq:motion}
  \ta s(\vec e, \vec q, \gamma) = \left(\frac1{\gamma-\gamma^{-1}}\vec F_{\perp}, \vec e, F_\parallel\right)
\end{math}
where $\vec F=\frac em\left(\beta \vec e\times\vec B + \vec E\right)$ and
$F_{\parallel} = (\vec F\cdot\vec e)$, $\vec F_\perp=\vec F-F_\parallel \vec e$.
Each slice  contains a reference to  the geometry and the parameters
of the encompassing beamline element (such as drift, bend, and quad).
 If we are  now to find the parameters of a particle
for a given time or arc  length,   we can do so by a two-step process:
%\begin{enumerate}
%\item 
(1) The slice $n$ in which the particle is at that $t$ or $s$ is
found by a binary search over the stored positions on the slices;
%\item
(2) The particle is ``micro-tracked'' into the slice by an amount of $t-t_n$,
using the initial conditions stored at the slice's entrance.
%\end{enumerate}
By this combination of storing and calculating trajectories, we
are able to efficiently refer to the history of particles in
computations to any desired precision.
\ifthenelse{\boolean{shortversion}}{}{
For efficiency   reasons,  the routine   doing the  micro-tracking  is
controlled by a bit-field telling it which parameters to output. E.~g.,
the local acceleration  of a particle  is not  needed for finding  the
retarded position of a  particle, although it  is needed for the field
calculation.

The usual phasespace-coordinates are regained by, for a given instant,
finding  $t_0$ for  which   $|\vec p(t) -  \vec   p_0(t+t_0)|$ becomes
minimal.  $\vec  p(t) -  \vec   p_0(t+t_0)$ then is  decomposed   with
respect  to  the  local  dreibein   tangential  to  $p_0(t+t_0)$  whose
$(x,l)$-plane coincides with the plane of motion in the last preceding
bend.
}
\subsection{Field Calculation}
Extended
charged distributions are used to model the bunch generating the field. One can not
use point particles for this purpose: as it is known from the theory
of synchrotron radiation, a point particle on a trajectory of bending
radius $\rho$ generates a radiation pulse of time width given by the critical frequency: $\tau \approx
\frac1{\omega_c}\approx \rho\gamma^{-3}$.  If a one-dimensional bunch of length
$\sigma_z$ was to be modeled by point particles, one would need
$N\approx\frac{\sigma_z}\rho\gamma^3$ particles and $O(N^2)$
operations when using a point-to-point algorithm. Instead, \progname{} tracks the
centroids of smeared-out bunchlets; a smooth field is generated by
numerically integrating over the distribution \cite{Dohlus:1996wr,Dohlus:1997am} and summing up the fields generated by the bunchlets.

1-  or 2-dimensional bunches  are smeared  along their trajectory or
along   their  trajectory and  perpendicular  to the  plane of motion,
respectively.  For  the time  being,  one  plane of  motion  has to be
specified; in this regard, \progname{}  is not fully 3-dimensional yet
(a   new   version     of  the     field   calculation   lifts    this
restriction\cite{Dohlus:1999xx}).
The charge distribution is Gaussian in both directions.

% copy from EPAC paper

\subsection{Shielding}

The fields generated by the bunch can change substantially when
non-trivial boundary conditions are present.  \progname{} can handle a
very limited case of shielding, namely that of infinitely extended
conductive parallel plates. The shielding effects are incorporated by
summing up the fields due to the image charges of the generating
bunchlets up to a certain user-defined cut-off distance.  As the
trajectories of the test particles cannot cross the trajectories of
the images charges, no singularity will be encountered.  Therefore,
1-dimensional bunchlets can be used for the image charges,
substantially saving CPU time.

%Another case of shielding is handled by checking for the visibility of
%the  retarded point  at  the  sampling point:  a transversally  narrow
%vacuum chamber   may intersect the   sagitta of  the bend   arc,  thus
%inhibiting  radiation   interaction.   \progname's field   calculation
%routines  optionally call a routine checking  this.  This concept also
%allows for the treatment of injector guns.

\subsection{Perturbative Tracking}

In most practical instances, one can assume that the changes of a
trajectory due to self-forces is small as compared to the
characteristic dimensions (bending radii, focal lengths) of the
system. Thus, the fields generated by a bunch not affected by its own
field will not differ too much from the fields generated by a bunch
traveling under the influence of its self-generated fields.
Consequently, we can find an approximate solution of the problem by
(1) tracking a bunch of particles through the beamline, considering
only the external guiding fields and (2) tracking a second bunch with
identical initial conditions, this by solving the discretized version
of \ref{eq:motion} in the presence of the fields which can be
calculated by referring to the unperturbed trajectory. The tracked
bunch need not be an exact copy of the generating bunch. When
calculating emittance dilutions for FEL applications, one is
interested in the slice emittance of a slice about the length of the
FEL slippage length. For this purposes, a sampling bunch can be used.
It consists of point particles, can have initial parameters different
from the generating bunch and does not contribute to the field
calculation, but samples only the fields generated by the
higher-dimensional generating bunchlets.

% For long version: singularities / 1d bunclets

\subsection{Self-Consistent Tracking}

The approach described above will not  work if the self-interaction of
the  bunch will  generate significant deviations  from the unperturbed
trajectory --  e.~g., a bunch in a  compressor chicane might collect a
correlated longitudinal  energy spread of  the same order of magnitude
as the one induced before the chicane. Then, the bunch length will not
agree with  the  one generated by  unperturbed tracking; consequently,
the calculated CSR fields and resulting phase space distributions will
be  incorrect. As  the problem   is  causal,   we can  approximate   a
self-consistent solution in the  following way: Two bunches  $B_0$ and
$B_1$ with identical  initial  conditions are  created.  In turns, the
bunches   are tracked through the   sequence of slices. While tracking
bunch $B_i$ through slice $n$, it is kicked by the fields generated by
$B_{1-i}$.  The deviation of  the  trajectories of $B_{0,1}$ gives  an
estimate  of the deviation  from the self-consistent  solution; it can
diminished by  refining the  discretization  of the beamline   and the
bunch population. When using a separate sampling bunch with this approach,
the average of the fields generated by $B_{0,1}$ is applied after they have
been tracked.

\ifthenelse{\boolean{shortversion}}{}{
\subsection{Bunches}

Bunches consist of  particles (0-dimensional charges) or bunchlets (1-
or 2-dimensional charge distributions).   Particles are used to sample
field distributions.

Three methods of beam population are implemented in
\progname: The bunchlets can either be set on a regular lattice in
six-dimensional phasespace, they can be distributed quasi-randomly or
pseudo-randomly.  All three methods are based on a general module for
generating distributions with a given correlation matrix $\sigma_{ik}
= \left\langle x_ix_k\right\rangle$.The Cholesky decomposition $C$,
where $CC^T=\Sigma$, is used to transform a normal-distributed vector
sequence into one with given $M$.  The vector sequence can be
generated by
\begin{enumerate}
\item enumerating a $d$-dimensional parallelepiped lattice
\item enumerating a $d$-dimensional Sobol sequence
\item generating $d$-tuples of Gaussian distributed pseudo-random numbers
\end{enumerate}
Method (1) is used to generate the generating  bunch, while (2) or (3)
are used for the sampling bunch. 
\ifthenelse{\boolean{shortversion}}{}{
In  the \progname{} input  file, the  correlation   matrix is  not given
directly, but in its traditional form (assuming vanishing correlations
between  transversal  planes   and   longitudinal  plane), namely   by
specifying    Twiss   parameters   $\alpha,\beta,\epsilon$  for    the
transversal    plane   and    $\delta_{incoherent}, \delta_{coherent},
\sigma_z$ for the longitudinal plane.}
}
\ifthenelse{\boolean{shortversion}}{}{
\subsection{Output Data}
\progname{} can provide  the complete information  gathered in a run for
further processing: the fields acting on each point and the phasespace
co-ordinates of each particle.  However,  this huge amount of data can
be suppressed.  \progname{} also calculates  the more useful  collective
quantities (rms  values, average  values,  Twiss  parameters, transfer
matrices, non-linear transfer matrices to  2nd order, chromaticities). 
The emittance is the usual  statistical  emittance; also, the area  of
the convex hull of the  n-$\sigma$ particles is calculated.  The  data
is written in ASCII  format into a  single output file, from which its
parts (e.g., Twiss parameters vs.  beamline position) can be extracted
with an included tool  for easy post-processing. For identification and
debugging   purposes, the complete parameters   of  the run, including
version information on all modules, are written to the output file.
}

\ifthenelse{\boolean{shortversion}}{}{
\subsection{The Code}
\progname{} is written in FORTRAN77 (field calculation) and
ANSI C++ (tracking, setup and evaluation). It currently
comprises about $10\,000$ lines of sources text. Its object-oriented
approach allows for easy augmentation by new types of elements.
A symbolic input language with the possibility to define
beamline-valued functions makes it easy to check beamline design alterations.
}

\section{Application Examples}
\subsection{The CTF2 Bunch Compressor}
\progname{} was  used to simulate  the emittance dilution in  the CLIC
Test  Facility Bunch Compressor\cite{D'Amico:1998vm},  which  has been measured\cite{CLIC}.
\begin{figure}
\begin{center}
\includegraphics[height=0.5\linewidth]{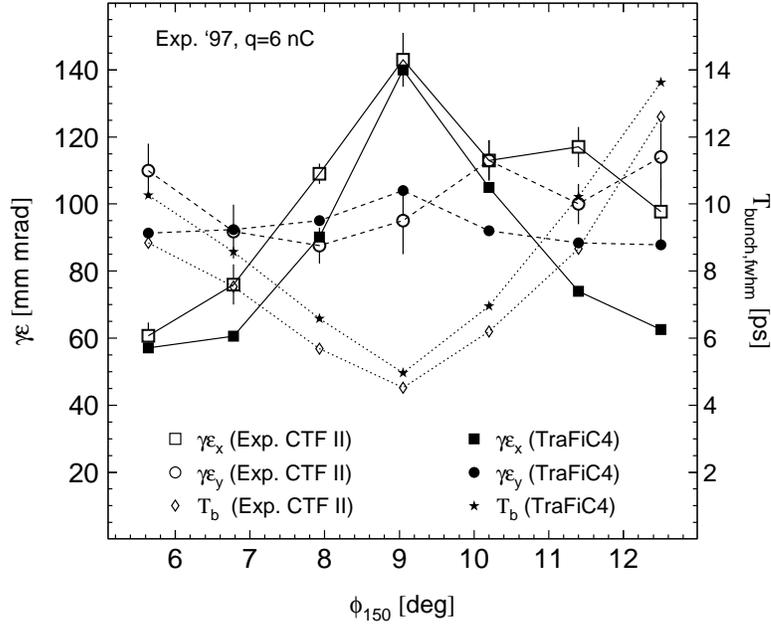}
\end{center}
\caption{Emittance measurements and simulation results for the CLIC Test Facility}
\label{fig:ctf2}
\end{figure}
Figure  \ref{fig:ctf2}\cite{CLIC,Groening}     shows the     results     for
$\epsilon_{x,y}$ and $\sigma_z$.  There is some underestimation of the
emittance growth for bending angles left of the peak.  A cause of this
might be the overestimation of the bunch length, which might stem from
a non-Gaussian bunch in the experiment. On the other side of the peak,
the  experiment shows  a  strongly deviating  behavior in $\epsilon_x$ and
$\epsilon_y$.  As  $\epsilon_y$  also increases, one  may  conclude that other
sources than CSR-induced   dispersive mismatch of  emittance might  be
responsible for that,  such as  the  proximity of the vacuum   chamber
wall. However, the agreement is quite reasonable, and the signature of
the experimental result is clearly reproduced.

\subsection{LCLS Dogleg}
\begin{figure}
\begin{center}
\includegraphics[width=0.85\linewidth]{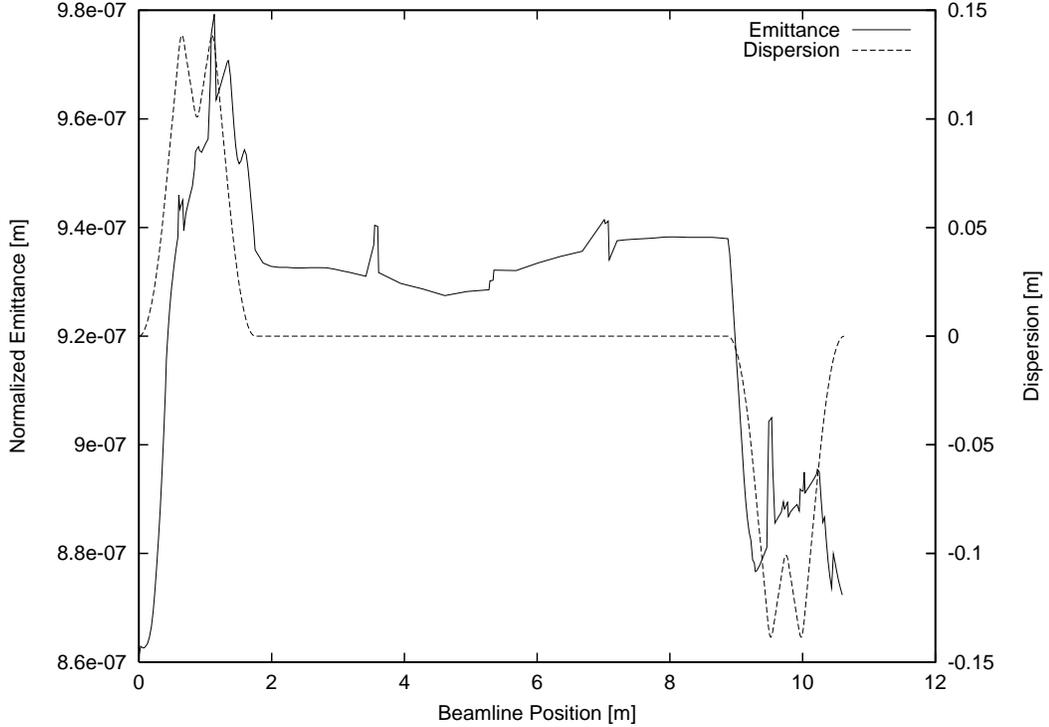}
\end{center}
\caption{Emittance growth compensation along the proposed LCLS injector dogleg}
\label{fig:lcls}
\end{figure}

Emittance growth induced by CSR and dispersion mismatch is a highly
correlated process. This means that it can, in principle, be undone: one can
untangle the disturbed transverse phasespace by applying to each particle
the opposite dispersive kick it suffered at an earlier
stage\cite{Emma:1997hj}. For an example, consider the optics for a proposed
dogleg injector layout for the LCLS\cite{Emma:Private}. It is used to
transport a 150MeV beam to a parallel offset tunnel through two bends of
$\pm38\deg$. The bends are realised by four bending magnets, which have
signature $+{}_1+-{}_2-$, and some quadrupole chosen such that the transfer
matrix $T_{x,12}=1$. \ref{fig:lcls} shows the transverse projected emittance growth 
for this setup as calculated by \progname; the $8\%$ growth after the first bends is
almost completely canceled between the last ones.
This is possible because the bunch retains its length, so the fields
have the same behavior in each bend. In bunch compression chicanes,
however, one has to chose a different scheme. One possibility is
to use several chicanes with an $1$ or $-1$ transfer matrix between them 
and to scale magnet lengths and dispersion according to the
expected field strengths.

\subsection{TTF Bunch Compressor}
\begin{figure}
\begin{center}
\includegraphics[width=0.85\linewidth]{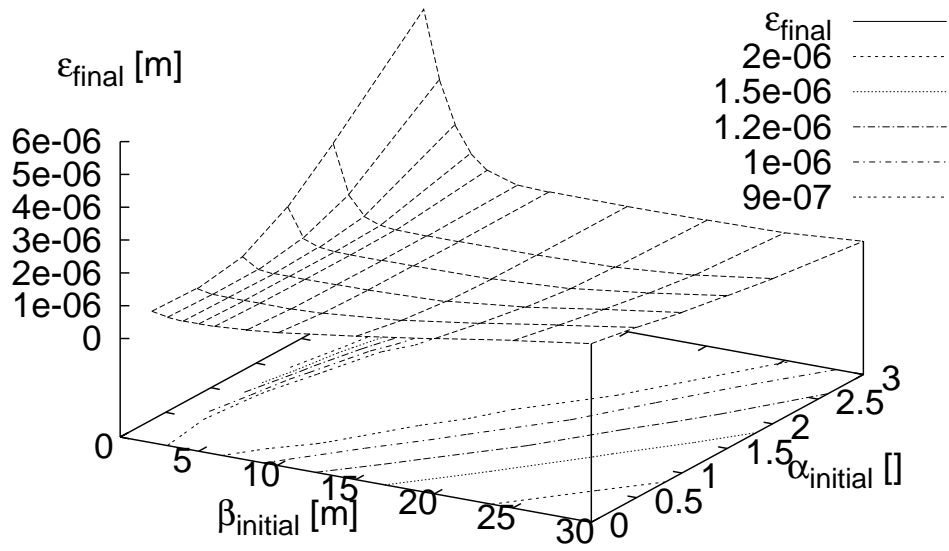}
\end{center}
\caption{Slice emittance growth in the TESLA Test Facility BCII chicane
as a function of initial Twiss parameters (from \cite{Dohlus:1999wj})}
\label{fig:paramscan}
\end{figure}

Even  in the absence of adjustable  optical elements there is room for
optimization  in  terms of  emittance.  The  TESLA Test Facility Bunch
Compressor II\cite{Geitz:1999wm} comprises four bending magnets and no quadrupoles within
the   dispersive  section. Coherent synchrotron Radiation
is a serious issue given its  parameter set \cite{Dohlus:1997zi,Dohlus:1997zj,Dohlus:1998vm}
However,   one  can  use  three quadrupoles
upstream of the chicane to adjust the initial Twiss parameters. 
%Figure
%\ref{fig:nonlinear} shows   the development  of  the transversal slice
%phase space as  the bunch travels to  the chicane.  
As the beam distortion exhibits a non-linear behavior, this can be utilized to cancel some of the induced kicks by an  appropriate choice of transverse beam
sizes and  divergences.   (A  purely linear  behavior would
leave the  emittance growth invariant under symplectic transformations
of the initial phasespace).
Figure \ref{fig:paramscan}  shows the results of a parameter
scan of the initial Twiss parameters; the differences in slice emittance
strongly suggest operating the compressor near the ``sweet spot'' around
$\alpha=1.2$, $\beta=15\textrm{m}$

%\cite{Talman:1986is}
%\cite{Lee:1988xx}
%\cite{Dohlus:1997am}
%\cite{Dohlus:1997zj}
%\cite{Dohlus:1998vm}
%\cite{Dohlus:1999wj}
%\cite{Dohlus:1996wr}
%\cite{Dohlus:1997zi}
%\cite{Derbenev:1995eg}
%\cite{Derbenev:1996ub}

\end{document}